\documentclass[aps,onecolumn,showpacs]{revtex4}
\usepackage{dcolumn}
\usepackage{tikz}
\usepackage{graphicx}
\usepackage{amsmath}
\usepackage{amsfonts}
\usepackage{amssymb}
\usepackage{psfrag}
\usepackage{wrapfig}
\usepackage{subfigure}
\usepackage{makeidx}
\usepackage{bm}
\usepackage{epsf}
\usepackage{pgflibraryarrows}
\usepackage{pgflibrarysnakes}

\begin{document}

\title{Defocusing complex short pulse equation and its multi-dark soliton solution}
\author{Bao-Feng Feng$^{1}$}
\author{Liming Ling$^{2}$}
\email{Corresponding_author: linglm@scut.edu.cn,baofeng.feng@utrgv.edu,  znzhu@sjtu.edu.cn}
\author{Zuonong Zhu$^{3}$}
\affiliation{$^1$ School of Mathematical and Statistical Sciences, The University of
Texas Rio Grande Valley, Edinburg Texas, 78539, USA}
\affiliation{$^2$School of Mathematics, South China University of
Technology, Guangzhou 510640, China}
\affiliation{$^3$Department of Mathematics, Shanghai Jiaotong University, Shanghai, China}

\date{April 19, 2016}

\begin{abstract}
In this paper,  we propose a complex short pulse equation of both focusing and defocusing types, which governs
the propagation of ultra-short pulses in nonlinear optical fibers. It can be viewed as an analogue of the nonlinear Schr\"odinger (NLS) equation in the ultra-short pulse regime. Furthermore, we construct the multi-dark soliton solution for the defocusing complex short pulse equation through the Darboux transformation and reciprocal (hodograph) transformation. One- and two-dark soliton solutions are given explicitly, whose properties and dynamics are analyzed and illustrated.
\newline
\end{abstract}

\maketitle



\section{Introduction}
It is well known that the nonlinear Schr\"odinger (NLS) equation, which describes the evolution of slowly varying wave packets waves in weakly nonlinear dispersive media
 under quasi-monochromatic assumption, has been very successful in many applications such as nonlinear optics and water waves
 \cite{Yarivbook,Kodamabook,Agrawalbook,Ablowitzbook}. However, as the width of optical pulses is in the order of femtosecond ($10^{-15}$ s),
 the spectrum of this ultra-short pulses is approximately of the order $10^{15} s^{-1}$, the monochromatic assumption to derive the NLS equation
 is not valid anymore \cite{Rothenberg}. Description of ultra-short processes requires a modification of standard slow varying envelope models.
This is the motivation for the study of the short pulse equation, the complex short pulse equation and their coupled models.

In 2004, Sch\"{a}fer and Wayne derived a short pulse (SP) equation \cite%
{SPE_Org}
\begin{equation}
u_{xt}=u+\frac{1}{6}\left( u^{3}\right) _{xx}\,,  \label{SPE}
\end{equation}%
to describe the propagation of ultra-short optical pulses in nonlinear media
\cite{SPE_CJSW}. Here $u=u(x,t)$ is a real-valued function, representing
the magnitude of the electric field, the subscripts $t$ and $x$ denote
partial differentiation. The SP equation has been shown to be completely
integrable \cite{Sakovich,Brunelli1,Brunelli2}, whose periodic and
soliton solutions of the SP equation were found in \cite%
{Sakovich2,Kuetche,Parkes,Matsuno_SPE,Matsuno_SPEreview}. 

Similar to the NLS equation, it is known that the complex-valued function has advantages in describing optical waves which have both the amplitude and phase information \cite{Yarivbook}. Following this spirit, one of the authors recently proposed a complex short pulse (CSP) equation \cite{Feng_ComplexSPE,FengShen_ComplexSPE}
\begin{equation}
q_{xt}+q+\frac{1}{2}\left( |q|^{2}q_{x}\right) _{x}=0.  \label{CSP}
\end{equation}%
In contrast with no
physical interpretation of the one-soliton solution to the SP equation (\ref{SPE}), the
one-soliton solution of the CSP equation (\ref{CSP}) is an envelope soliton with a few
optical cycles. 

The CSP equation can be viewed as an analogue of the NLS equation
in the ultra-short pulse regime when the width of optical pulse is of the order $10^{-15} s$.
The NLS equation has the focusing and defocusing cases, which admits the bright and dark type soliton solutions, respectively. As a matter of fact, the dark soliton in optical fibers was predicted in 1973 \cite{Hasekawa73}, and was observed experimentally in 1988 \cite{Darkexp1,Darkexp2}, a decade earlier than the observation of the bright soliton. Therefore
it is natural that the CSP equation can also have the focusing and defocusing type, which may be proposed as
\begin{equation}
q_{xt}+q+\frac{1}{2} \sigma \left( |q|^{2}q_{x}\right) _{x}=0\,,  \label{gCSP}
\end{equation}
where $\sigma=1$ represents the focusing case, and $\sigma=-1$ stands for the defocusing case.
It turns out that this is indeed the case as shown in the subsequent section.
Same as the focusing CSP equation discussed in \cite%
{Feng_ComplexSPE,FengShen_ComplexSPE}, the defocusing CSP equation can also
occur in nonlinear optics when ultra-short pulses propagate in a nonlinear
media of defocusing type.

The remainder of the present paper is organized as follows. In section II,
the CSP equation of both the focusing and defocusing types is derived
from the context of nonlinear optics based on Maxwell's equations. Then, based on the reciprocal link between the
defocusing CSP equation and the complex coupled dispersionless (CCD) equation,
the Darboux transformation to the CCD equation is derived to give a general solitonic formula to the
defocusing CSP equation in section III. We continue to derive explicit formulas for one- and multi-dark soliton
solutions to the defocusing CSP equation by a limiting process in section IV.
The one- and two-dark soliton solution is analyzed in details, which can be classified into smoothed, cusponed and looped ones depending on the parameters.
The paper is concluded by some comments and remarks in section V.
\section{Derivation of the focusing and defocusing complex short pulse equation}
The starting point to derive the CSP equation is the same as the one for the NLS equation
\cite{Kodamabook,Agrawalbook,Ablowitzbook}, which is the celebrating Maxwell's equations
\begin{equation} \label{Maxwell}
\nabla\times \mathbf{E} =- \frac{\partial \mathbf{B}}{\partial t}\,, \quad
\nabla\times \mathbf{H} =- \frac{\partial \mathbf{D}}{\partial t}\,,
\end{equation}
where $\mathbf{E}$ and $\mathbf{H}$ are electric and magnetic field vectors, and $\mathbf{D}$ and $\mathbf{B}$ are
corresponding electric and magnetic flux densities. The relations between $\mathbf{D}$, $\mathbf{B}$ and
$\mathbf{E}$, $\mathbf{H}$ are called the constitutive relations given by
\begin{equation} \label{Constitutive}
\mathbf{D}= \epsilon \mathbf{E}\,, \quad \mathbf{B}= \mu \mathbf{H}\,,
\end{equation}
where $\epsilon$ is the permittivity, $\mu$ is the permeability. In vacuum, $c^2=1/(\epsilon_0 \mu_0)$ with $c$ the velocity of light in vacuum. In the frequency-dependent media,
\begin{equation} \label{Constitutive2}
\mathbf{D}= \epsilon \star \mathbf{E}\,, \quad \mathbf{B}= \mu \star\mathbf{H}\,, \quad \mathbf{D}=\mathbf{E} +\mathbf{P}\,,
\end{equation}
where $\star$ means the convolution, and $\mathbf{P}$ is the electric induced polarization. By eliminating $\mathbf{B}$ and $\mathbf{D}$, the following wave equation follows
\begin{equation}  \label{E-wave-equation}
\nabla^2 \mathbf{E} -\frac{1}{c^2} \mathbf{E}_{tt} = \mu_0 \mathbf{P}_{tt}\,,
\end{equation}
which describes light propagation in optical fibers.
If we assume the local medium response and only the third-order nonlinear effects governed by $\chi^{(3)}$, the induced
polarization consists of linear and nonlinear parts, $\mathbf{P} ( \mathbf{r},t)=\mathbf{P}%
_{L} ( \mathbf{r},t)+\mathbf{P}_{NL} ( \mathbf{r},t)$, where the linear part
\begin{equation}  \label{P_L}
\mathbf{P}_{L} ( \mathbf{r},t)=\epsilon_0 \int_{-\infty}^{\infty} \chi^{(1)}
(t-t^{\prime })\cdot \mathbf{E} ( \mathbf{r},t^{\prime })\,dt^{\prime }\,,
\end{equation}
and the nonlinear part
\begin{equation}  \label{P_NL}
\mathbf{P}_{NL}( \mathbf{r},t)=\epsilon_0 \int_{-\infty}^{\infty} \chi^{(3)}
(t-t_1,t-t_2,t-t_3)\times \mathbf{E} ( \mathbf{r},t_1) \mathbf{E} ( \mathbf{r%
},t_2) \mathbf{E} ( \mathbf{r},t_3)\,dt_1dt_2dt_3\,.
\end{equation}
Here $\epsilon_0$ is the vacuum permittivity and $\chi^{(j)}$ is the $j$th-order susceptibility.
As discussed in \cite{Brabec}, the nonlinear response is due to the induced atomic dipole with a response time
of the order $1/\Delta$, where $\Delta=|\omega_{ik}-\omega_0|$. $\omega_{ik}$
represents the transition frequency from the
initial (usually ground) quantum state $i$ into some excited
state $k$, and $\omega_0$ is the central carrier frequency. Since the
typical transition frequency from the atomic ground
state to the lowest excited state significantly exceeds the
usual carrier frequency, $1/\Delta$ is typically less than 1 fs. Therefore, we can assume an instantaneous nonlinear
response in femtosecond regime.
Moreover, the nonlinear effects are relatively small in
silica fibers, $\mathbf{P}_{NL}$ can be treated as a small perturbation.
Therefore, we first consider Eq. (\ref{E-wave-equation}) with $\mathbf{P}_{NL}=0$. %
Furthermore, we restrict ourselves to the case that the optical pulse
maintains its polarization along the optical fiber, and the transverse
diffraction term
can be neglected. In this case,
the electric field can be considered to be one-dimensional and expressed as
\begin{equation}  \label{E-field}
\mathbf{E} = \frac 12 \mathbf{e_1} \left(E(z,t)+c.c. \right)\,,
\end{equation}
where $\mathbf{e_1}$ is a unit vector in the direction of the polarization,
$E(z,t)$ is the complex-valued function, and $c.c.$ stands for the complex
conjugate.
Under this case, it is useful to transform Eq. (\ref{E-wave-equation}) into the frequency domain, which reads
\begin{equation}  \label{Waveequation-Fourier}
\tilde {E}_{zz} (z,\omega) + \epsilon(\omega) \frac{\omega^2}{c^2} \tilde {E}
(z,\omega)=0\,,
\end{equation}
where $\tilde {E} (z,\omega)$ is the Fourier transform of $E(z,t)$ defined
as
\begin{equation}  \label{E_Fourier}
\tilde E (z,\omega)=\int_{-\infty}^{\infty} {\ E} (z,t) e^{\mathrm{i} \omega
t}\,dt\,.
\end{equation}
The frequency-dependent dielectric constant occurring in Eq. (\ref{E_Fourier}) is defined as
\begin{equation}  \label{Dielectric}
\epsilon(\omega)=1+ \tilde \chi^{(1)} (\omega)\,,
\end{equation}
where $\tilde \chi^{(1)} (\omega)$ is the Fourier transform of $\chi^{(1)}(t)$. Up to now, the consideration
is exactly the same as the one for deriving the NLS equation. To derive the NLS equation, the optical field is assumed to be quasi-monochromatic, i.e., the pulse spectrum, centered as $\omega_0$, is assumed to have a spectral width $\Delta \omega$ such that $\Delta \omega/ \omega_0 \ll 1$. Under this assumption, the NLS equation can be derived to govern the slowly varying envelop of optical wave packet in weakly nonlinear dispersive media.
However, when the width of optical pulse is in the order of femtosecond ($10^{-15}$ s), the monochromatic assumption to derive the NLS equation is not valid anymore. We need to construct a suitable fit to the frequency-dependent dielectric constant $\epsilon(\omega)$ in the desired spectral range.
More specifically, for the frequency-dependent dielectric constant $\epsilon (\omega )=1+\tilde{\chi}^{(1)}(\omega )$, we
assume $\tilde{\chi}^{(1)}$ can be approximated by
\begin{equation}
\tilde{\chi}^{(1)}=\tilde{\chi}_{0}^{(1)}\mp \tilde{\chi}_{2}^{(1)}\lambda
^{2}\,,\quad \tilde{\chi}_{2}^{(1)}>0\,.
\end{equation}
As discussed subsequently, the negative sign represents the focusing media with anomalous group velocity dispersion (GVD), and the positive sign stands for the defocusing media with normal GVD.

Next we proceed to the consideration of the nonlinear effect. Assuming the
nonlinear response is instantaneous so that $P_{NL}$ is given by $%
P_{NL}(z,t)= \epsilon_0 \epsilon_{NL} E(z,t)$ \cite{Agrawalbook} where the
nonlinear contribution to the dielectric constant is defined as
\begin{equation}  \label{Epsnl}
\epsilon_{NL}= \frac 34 \chi^{(3)}_{xxxx} |E(z,t)|^2\,.
\end{equation}
Therefore, the Helmholtz equation can be modified as
\begin{equation}  \label{Helmholtz}
\tilde {E}_{zz} (z,\omega) + \tilde \epsilon(\omega) \frac{\omega^2}{c^2}
\tilde {E} (z,\omega)=0\,,
\end{equation}
where
\begin{equation}  \label{Modified_Dielectric}
\tilde \epsilon(\omega)=1+\tilde{\chi}_{0}^{(1)}\mp \tilde{\chi}_{2}^{(1)}\lambda
^{2}+ \epsilon_{NL}\,.
\end{equation}
In summary, Eq. (\ref{Helmholtz}) with Kerr cubic nonlinearity reads
\begin{equation}
\tilde{E}_{zz}+\frac{1+\tilde{\chi}_{0}^{(1)}}{c^{2}}\omega ^{2}\tilde{E}\mp
(2\pi )^{2}\tilde{\chi}_{2}^{(1)}\tilde{E}+\epsilon _{NL}\frac{\omega ^{2}}{%
c^{2}}\tilde{E}=0\,.  \label{Wave_Fourier2}
\end{equation}%
By applying the inverse Fourier transform to Eq. (\ref{Wave_Fourier2}), the nonlinear wave equation in physical domain  is
\begin{equation}
E_{zz}-\frac{1}{c_{1}^{2}}E_{tt}=\pm \frac{1}{c_{2}^{2}}E+\frac{3}{4}\chi
_{xxxx}^{(3)}\left( |E|^{2}E\right) _{tt}\,,  \label{Wave_nonlinear1}
\end{equation}%
where
\begin{equation}
c_1=\frac{c}{\sqrt{1+\tilde{\chi}_{0}^{(1)}}}\,,\quad c_2=\frac{1}{2\pi \sqrt{\tilde{\chi}_{2}^{(1)}}}\,.
\end{equation}%
Furthermore, by using the normalized independent variables $z \to c_2 z$, $t \to c_2/c_1$ and normalized field
$E \to (3c_1^2 \chi_{xxxx}^{(3)}/4)^{-1/2} E$, we obtain the normalized wave equation
\begin{equation}
E_{zz}-E_{tt}=\pm E+ \left( |E|^{2}E\right) _{tt}\,.  \label{Wave_nonlinear1b}
\end{equation}%
Next, we focus on only a right-moving wave packet and
assume a multiple scales \textrm{ansatz}
\begin{equation}  \label{E_ansatz}
E(z,t)=\epsilon E_0(\tau, z_1, z_2, \cdots)+ \epsilon^2 E_1(\tau, z_1, z_2,
\cdots) + \cdots\,,
\end{equation}
where $\epsilon$ is a small parameter, $\tau$ and $z_n$ are the scaled
variables defined by
\begin{equation}  \label{Multiple_ansatz}
\tau= \frac{t-{x}}{\epsilon}, \quad z_n=\epsilon^n z\,.
\end{equation}
Substituting (\ref{E_ansatz}) with (\ref{Multiple_ansatz}) into (\ref%
{Wave_nonlinear1}), we obtain the following partial differential equation
for $E_0$ at the order $O(\epsilon)$:
\begin{equation}  \label{Wave_nonlinear2}
- 2 \frac{\partial^2 E_0}{\partial{\tau}\partial{z_1}} = \pm  E_0 + 2\frac{\partial}{\partial{\tau}}
\left(|E_0|^2 \frac{\partial E_0}{\partial {\tau}} \right)\,.
\end{equation}
Here the term $E^2_0 E_{0,\tau}$ is ignored but it is validated subsequently.
Finally a general complex short pulse equation can be obtained
\begin{equation}
q_{xt}\pm q+\frac{1}{2}\left( |q|^{2}q_{x}\right) _{x}=0\,.  \label{gCSP2}
\end{equation}%
by the scale transformations
\begin{equation}  \label{Scaling}
x= \frac{1}{\sqrt{2}} \tau, \quad t=\frac{1}{\sqrt{2}} z_1, \quad q = \sqrt{2} E_0\,.
\end{equation}
It is obvious that Eq. (\ref{gCSP2}) with positive sign is the same as Eq. (\ref{gCSP}), while
Eq. (\ref{gCSP2}) with negative sign is equivalent to Eq. (\ref{gCSP}) by a conversion of time $t \rightarrow -t$. Consequently, we derive the CSP equation of both the focusing and defocusing types. We should pointed out that there are typos in the scaling transformations in \cite{Feng_ComplexSPE}.

To validate the approximation, we compare the solutions to Maxwell equations and with the ones to the CSP equation. As a matter of fact, solitary wave solutions with a few cycles derived directly from the Maxwell equations under the assumption of the Kramers-Kronig relation holds have been investigated in the literature \cite{Skobelev,Kim,Amir,Amir2}.
Here we mainly refer to the results in \cite{Kim} and consider the normalized equation (\ref{Wave_nonlinear1b}) with positive sign. We assume an envelop solitary wave solution is of the form
\begin{equation}  \label{Ansatz_Maxwell}
E(z,t)=A(\xi) e^{\mathrm{i} \phi(z,t)} \,,
\end{equation}
with $\xi=z-vt$, $\phi(z,t)=\omega(t-vz) + F(\xi)$.
Inserting this ansatz into Eq. (\ref{Wave_nonlinear1b}), one obtains the set of equations
\begin{equation}\label{Max1}
A_{zz}-A_{tt}-A(\phi_z^2 -\phi_t^2) -A -(6AA_t^2+3A^2A_{tt}-A^3\phi_t^2)=0\,,
\end{equation}
\begin{equation}\label{Max2}
2A_{z} \phi_z+ A\phi_{zz} -(2A_t\phi_t + A\phi_{tt}) - (6A^2A_t \phi_t+A^3\phi_{tt})=0\,,
\end{equation}
 \begin{equation}\label{Max3}
[1-v^2 A^2/(1-v^2)] A F_{\xi\xi} + 2[1-3v^2 A^2/(1-v^2)] A_{\xi} F_{\xi} + 6 v\omega A^2A_{\xi}/(1-v^2) =0\,.
\end{equation}
 By introducing normalized amplitude $a=vA/\sqrt{1-v^2}$, we obtain
\begin{equation}\label{Max4}
F(\xi)=-\frac{\omega }{2v} \int_{-\infty}^{\xi} \frac{a^2(3-2a^2)}{(1-a^2)} d\,\xi'\,
\end{equation}
by integrating Eq. (\ref{Max3}) once. Further, inserting $\phi_t=\omega - vF_\xi$ and $\phi_z=-\omega v + F_\xi$ into Eq. (\ref{Max1}), one obtains a second-order differential equation
\begin{equation}\label{Max5a}
a_{\xi \xi} - \frac{6aa^2_\xi}{1-3a^2} -\frac{\omega^2a}{1-3a^2} \left(\frac{\delta^2}{v^2} - \frac{a^2(4(1-a^2)^2-a^2)}{4v^2(1-a^2)^3}\right) =0\,,
\end{equation}
where $\delta^2 = v^2/(\omega^2(1-v^2)) -v^2$.
Integrating once and requiring $a, a_{\xi} \to 0$ at $\xi \to \pm \infty$, one arrives at
\begin{equation}\label{Max5}
a_\xi=\pm \frac{\omega \sqrt{1+\delta^2} a}{v(1-3a^2)(1-a^2)} \sqrt{\left(1-\frac 32 a^2\right) \left(\frac{1+4\delta^2+\sqrt{1-8\delta^2}}{4(1+\delta^2)} -a^2 \right) \left( \frac{1+4\delta^2-\sqrt{1-8\delta^2}}{4(1+\delta^2)} -a^2 \right)} \,.
\end{equation}
From (\ref{Max5}), one can easily show that a localized solution exist with amplitude
\begin{equation}\label{Max6}
a_{\text{max}}=\frac 12 \sqrt{\frac{1+4\delta^2+\sqrt{1-8\delta^2}}{1+\delta^2}}\,
\end{equation}
provided $\delta^2 \le 1/8$.

As mentioned in \cite{Kim}, in the case of $a_{\text{max}} \ll 1$ where slowly evolving
wave field approximation (SEWA) is valid, the solution to Eq. (\ref{Max5}) can be written as
\begin{equation}\label{Max7a}
\frac 92 \delta \sqrt{\delta^2-\frac{a^2}{2}} - \cosh^{-1} \left( \frac{\sqrt{2}\delta}{a}\right) = \pm \frac{\omega}{v} \delta \xi\,.
\end{equation}
Furthermore, when $\delta^2 \ll 1/8$ and the first term in  Eq. (\ref{Max7a}) can be neglected, we obtain the one-soliton solution to the NLS equation
\begin{equation}\label{Max8}
a_{\text{NLS}}=\sqrt{2}\delta {\mathrm sech} \left(\omega \delta \xi/v \right)  \,.
\end{equation}
Multiplying Eq. (\ref{Max7a}) by $2$ and taking cosh-function, we arrive at a localized solution to
the higher order nonlinear Schr\"odinger (HONLS) equation
by taking into account dispersions beyond group velocity dispersion (GVD)
\begin{equation}\label{Max7}
a_{\text{HONLS}}=\frac{2\sqrt{2}\delta}{9\delta^2+\sqrt{81\delta^2+4\cosh(2\omega \delta \xi/v)}}, \quad F=-\frac{3\omega}{2v} \int_{-\infty}^{\xi} a^2 d\,\xi'\,.
\end{equation}


\begin{figure}[tbh]
\centering
\includegraphics[height=60mm,width=80mm]{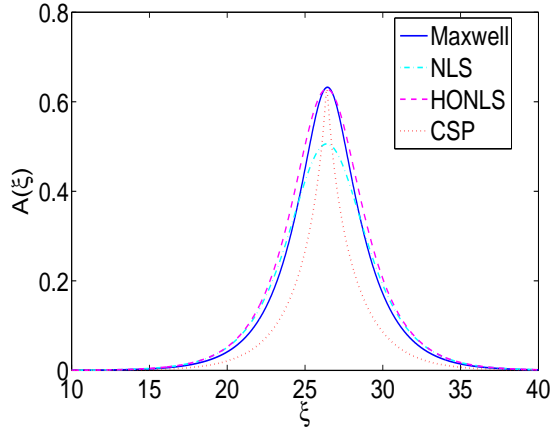} \hfil
\caption{(color online): Comparison of solitary solutions among Maxwell equations (solid blue line), focusing CSP equation (red dotted line), NLS (magenta dashed line) and the higher order NLS (cyan dash-dot line) equations.}
\label{fig1}
\end{figure}
In Fig. 1, we compare the solutions for Eq. (\ref{Wave_nonlinear1b}), Eq. (\ref{gCSP2}) with positive sign \cite{Feng_ComplexSPE,FengShen_ComplexSPE,LingFeng1}, the solution to the NLS equation and the higher order NLS equation (\ref{Max7}),  (\ref{Max7a}) for the parameters $v=1/2.25$, $\omega=1.0$.
Here, a classical Runge-Kutta method is used to integrate
Eq. (\ref{Max5}).
It can be observed that solution of the Maxwell equations lies in between the ones of the CSP equation and the higher order NLS equation.

For the defocusing case, through a similar procedure as the focusing case, we can obtain the following equations:
\begin{equation}\label{defocusing}
  \begin{split}
     &[a_{\xi}(3a^2-1)]_{\xi}+\frac{\omega^2}{4}\frac{a^4(-a^4+4(1-a^2)^2)-4C_0v^3(a^4+v^3C_0)}{v^2(a^2-1)^3a^3}-
     \frac{(v^2-1)^2\omega^2-v^2}{v^2(v^2-1)}a=0, \\
     &F(\xi)=\int_{-\infty}^{\xi}\left[\frac{\omega(2a^6-3a^4-2C_0v^3)}{2v(a^2-1)^2a^2}-C_2\right]\mathrm{d}\xi+C_2\xi.
  \end{split}
\end{equation}
where $C_0$ is an integration constant.
Integrating \eqref{defocusing} once, we arrive at
\begin{equation*}
  a_{\xi}=\pm\frac{\sqrt{G(a,v,\omega,C_0,C_1)}}{2v\sqrt{(1-v^2)}(3a^2-1)(a^2-1)a}
\end{equation*}
where $G(a,v,\omega,C_0,C_1)$ is a tenth order polynomial with respect to $a$(so we omit the explicit formula), $C_1$ is an integration constant.
In a special case, we can obtain the dark soliton solution by choosing the parameters $v=0.44$,$\omega=0.4799915339$, $C_0=-17.65400508$. Similarly, the dark soliton can be obtained by numerically solving \eqref{defocusing} via classical Runge-Kutta method. The result is compared with the one for the defocung CSP equation in Fig. 2. As is seen, a good agreement is achieved.
\begin{figure}[tbh]
\centering
\includegraphics[height=60mm,width=80mm]{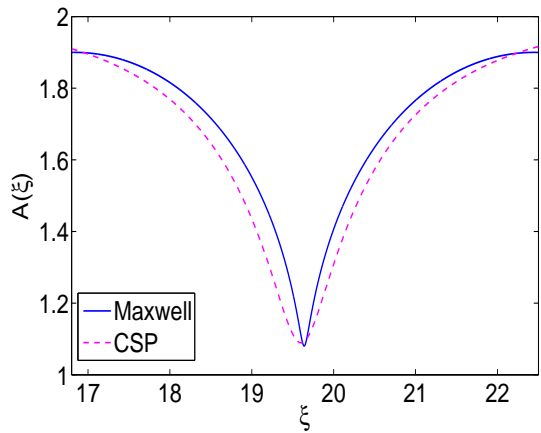} \hfil
\caption{(color online): Comparison of dark solitons between Maxwell equations (blue solid line) and the defocusing CSP equation (magenta dashed line).}
\label{fig2}
\end{figure}

Notice that the CSP equation (\ref{gCSP}) can be rewritten by
\begin{equation*}
(\sqrt{1+\sigma|q_x|^2})_t + \frac{1}{2} \sigma (|q|^2\sqrt{1+\sigma|q_x|^2})_x=0\,,
\end{equation*}
so that we can define a reciprocal (hodograph) transformation
\begin{equation}
\label{hodograph}
\mathrm{d}y=\rho^{-1} \mathrm{d}x - \frac{1}{2} \sigma \rho^{-1} {|q|^2} \mathrm{d%
}t,\,\, \mathrm{d}s=-\mathrm{d}t,
\end{equation}
where $\rho^{-1}=\sqrt{1+\sigma|q_x|^2}$. By doing so, the CSP equation (\ref{gCSP}) is converted into the following coupled equation
\begin{equation}  \label{CCDa}
q_{ys}=\rho q\,,
\end{equation}
\begin{equation}  \label{CCDb}
\rho_s + \frac{1}{2}\sigma (|q|^2)_y=0\,.
\end{equation}
We remark here that equations (\ref{CCDa})--(\ref{CCDb}) with $\sigma=1$ is the complex
coupled dispersionless (CCD) equation studied in \cite{KonnoKakuhata2}, while the
case of $\sigma=-1$ is the case which, for some reason, hasn't been studied in the literature.
\section{Darboux transformation and multi-dark soliton solution to the defocusing CSP
equation} \label{section3}
In the present section, we aim at finding the multi-dark soliton solution of the defocusing CSP equation
\begin{equation}
q_{xt}+q-\frac{1}{2}\left( |q|^{2}q_{x}\right) _{x}=0\,  \label{dCSP}
\end{equation}
via the Darboux transformation method. Firstly, it is noted that the CSP equation is invariant under the following
scaling transformations $q\rightarrow cq$, $\partial _{x}\rightarrow \frac{1}{c}\partial _{x}$ and
$\partial _{t}\rightarrow c\partial _{t}$. Thus, without loss of generosity, we can fix either the amplitude or the wavenumber of $q$.
Secondly, due to the fact that the CSP equation belongs to the
Wadati-Konno-Ichikawa (WKI) hierarchy, it is not feasible to construct the
Darboux transformation (DT) from the spectral problem of the CSP equation
directly. Instead, we can develop the DT for the CCD equation
which is linked to the CSP equation by the hodograph transformation (\ref{hodograph}).

In what follows, we present the Lax pair and the corresponding DT of the CCD equation (\ref{CCDa})--(\ref{CCDb}) with $\sigma=-1$. It can be easily shown that the compatibility condition $\Psi_{ys} = \Psi _{sy}$ of the following linear problems
\begin{equation}
\Psi _{y}= U(\rho ,q;\lambda )\Psi,
\label{ccd-laxa}
\end{equation}%
\begin{equation}
\Psi _{s}= V(q;\lambda )\Psi,
\label{ccd-laxb}
\end{equation}%
where
\begin{equation*}
U(q,\rho ;\lambda )=\lambda ^{-1}%
\begin{bmatrix}
-\mathrm{i}\rho & \bar{q}_{y} \\[8pt]
q_{y} & \mathrm{i}\rho%
\end{bmatrix}%
,\quad \,V(q;\lambda )=\frac{\mathrm{i}}{4}\lambda \sigma _{3}+\frac{\mathrm{i}}{%
2}Q,\quad \,Q=%
\begin{bmatrix}
0 & -\bar{q} \\[8pt]
q & 0%
\end{bmatrix}\,,
\end{equation*}%
with the overbar representing the complex conjugate and $\sigma _{3}$ being the third Pauli matrix, yields the defocusing CCD equation
\begin{equation}
q_{ys}= \rho q\,,
\label{dccda}
\end{equation}%
\begin{equation}
\rho _{s}-\frac{1}{2}(|q|^{2})_{y}= 0\,.
\label{dccdb}
\end{equation}%
Through the hodograph transformation
\begin{equation*}
\mathrm{d}x=\rho \mathrm{d}y+\frac{1}{2}|q|^{2}\mathrm{d}s,\,\,dt=-ds,
\end{equation*}%
one can obtain the defocusing CSP equation (\ref{dCSP}).
To obtain the soliton equation, we give the following Darboux matrix for the defocusing CSP equation
(\ref{dCSP}). We omitted the proof here, interested audience can refer to \cite{algebraic,Matveev,ling-dark,LingFeng1} for details.
\begin{equation}
T=I+\frac{\bar{\lambda}_{1}-\lambda _{1}}{\lambda -\bar{\lambda}_{1}}%
P_{1},\,\,P_{1}=\frac{|y_{1}\rangle \langle y_{1}|\sigma _{3}}{\langle
y_{1}|\sigma _{3}|y_{1}\rangle },\,\,\langle y_{1}|=|y_{1}\rangle ^{\dag
},\,\,|y_{1}\rangle =%
\begin{bmatrix}
\psi _{1}(y,s;\lambda _{1}) \\
\phi _{1}(y,s;\lambda _{1})%
\end{bmatrix}
\label{DT}
\end{equation}%
where $|y_1\rangle$ denotes the special solution for system \eqref{ccd-laxa}--\eqref{ccd-laxb} with $\lambda=\lambda_1$,
can convert system \eqref{ccd-laxa}--\eqref{ccd-laxb} into a new system
\begin{equation} \label{DTa}
\Psi \lbrack 1]_{y}= U(q,\rho ;\lambda )\Psi \lbrack 1]\,,
\end{equation}%
\begin{equation} \label{DTb}
\Psi \lbrack 1]_{s}= V(q;\lambda )\Psi \lbrack 1].
\end{equation}%
The B\"{a}cklund transformations between $(q[1],\rho \lbrack 1])$ and $%
(q,\rho )$ are given through
\begin{equation}
q[1] =q+\frac{(\bar{\lambda}_{1}-\lambda _{1})\bar{\psi}_{1}\phi _{1}}{%
\langle y_{1}|\sigma _{3}|y_{1}\rangle},
\label{backlunda}
\end{equation}
\begin{equation}
\rho \lbrack 1] =\rho -2\ln _{ys}\left( \frac{\langle y_{1}|\sigma
_{3}|y_{1}\rangle }{\bar{\lambda}_{1}-\lambda _{1}}\right)\,,
\label{backlundb}
\end{equation}
\begin{equation}
|q[1]|^{2} =|q|^{2}-4\ln _{ss}\left( \frac{\langle y_{1}|\sigma
_{3}|y_{1}\rangle }{\bar{\lambda}_{1}-\lambda _{1}}\right)\,.
\label{backlundc}
\end{equation}
Furthermore, we have the following $N$-fold Darboux matrix:
The $N$-fold Darboux matrix can be represented as
\begin{equation}
T_{N}=I+YM^{-1}D^{-1}Y^{\dag }\sigma _{3},
\label{DT_Nfold}
\end{equation}%
where
\begin{equation*}
\begin{split}
Y=& \left[ |y_{1}\rangle ,|y_{2}\rangle ,\cdots ,|y_{N}\rangle \right] , \\
M=& \left( \frac{\langle y_{i}|\sigma _{3}|y_{j}\rangle }{\bar{\lambda }%
_{i}-\lambda _{j}}\right) _{1\leq i,j\leq N}, \\
D=& \mathrm{diag}\left( \lambda -\bar{\lambda }_{1},\lambda -\bar{%
\lambda }_{2},\cdots ,\lambda -\bar{\lambda }_{N}\right)\,,
\end{split}%
\end{equation*}%
the vector $|y_i\rangle$ represents the special solution for system \eqref{ccd-laxa}-\eqref{ccd-laxb} with $\lambda=\lambda_i$, and the B\"{a}cklund transformations for $q[N]$ and $\rho[N]$ are
\begin{equation}
q[N] =q+\frac{\det (\widehat{M})}{\det (M)},
\label{gBT1}
\end{equation}%
\begin{equation}
\rho \lbrack N] =\rho -2\ln _{ys}\left( \det (M)\right)\,,
\label{gBT2}
\end{equation}%
\begin{equation}
|q[N]|^{2} =|q|^{2}-4\ln _{ss}\left( \det (M)\right)\,.
\label{gBT3}
\end{equation}%
where $\widehat{M}=%
\begin{bmatrix}
M & Y_{1}^{\dag } \\
-Y_{2} & 0%
\end{bmatrix}$, $Y_{k}$ represents the $k$-th row of matrix $Y$.
The proof can be given similar to the one in \cite{ling-dark}, which is omitted here.
Instead, we merely commented that the following identities associated with the matrix and determinant are used.
\begin{equation*}
\begin{split}
& \phi M^{-1}\psi ^{\dag}=%
\begin{vmatrix}
M & \psi ^{\dag } \\
-\phi & 0%
\end{vmatrix}/|M|, \\
& 1+\phi M^{-1}\psi ^{\dag }=%
\begin{vmatrix}
M & \psi ^{\dag } \\
-\phi & 1%
\end{vmatrix}%
/|M|=\frac{\det (M+\psi ^{\dag }\phi )}{\det (M)}.
\end{split}%
\end{equation*}
Here $M$ is a $N\times N$ matrix, $\phi $, $\psi $ are the $%
N\times 1$ matrix. Based on above $N$-fold Darboux transformation for the CCD system \eqref{CCDa}--\eqref{CCDb}, we
have the solitonic solution formula for the defocusing CSP equation \eqref{dCSP} in parametric form.
\begin{equation}
q[N] =q+\frac{\det(\widehat{M})}{\det (M)}\,,
\label{CSP-formula1}
\end{equation}
\begin{equation}
x =\int \rho \mathrm{d}y+\frac{1}{2}\int |q|^{2}\mathrm{d}s-2\ln _{s}\left(
\det (M)\right) ,\quad \,t=-s\,.
\label{CSP-formula2}
\end{equation}
\section{One- and multi-dark solutions to the defocusing CSP equation}
\label{section4}
In this section, we derive an explicit expression for the one- and multi-dark
soliton solution to the defocusing CSP equation through formulas \eqref{gBT1}--\eqref{gBT3} by a limit
technique.
\subsection{One-dark soliton solution}
We start with the seed solution
\begin{equation}
\rho \lbrack 0]=-\frac{\gamma }{2},\,\,q[0]=\frac{\beta }{2}\mathrm{e}^{%
\mathrm{i}\theta },\,\,\theta =y+\frac{\gamma }{2}s,\,\,\gamma >0.
\end{equation}
Introducing a gauge transformation with $\Psi =K\widehat{\Psi }$ with
\begin{equation*}
K=\mathrm{diag}\left( \mathrm{e}^{\frac{\mathrm{i}}{2}\theta },\mathrm{e}^{-%
\frac{\mathrm{i}}{2}\theta }\right)\, ,
\end{equation*}
we can solve the Lax pair equation \eqref{ccd-laxa}--\eqref{ccd-laxb} at $\lambda =\lambda
_{1} $, finding fundamental matrix solution as follows
\begin{equation*}
\Psi =KL_{1}M_{1}\,,
\end{equation*}
where
\begin{equation*}
\,\,L_{1}=%
\begin{bmatrix}
1 & 1 \\[8pt]
\frac{\beta }{\chi _{1}^{+}+\gamma } & \frac{\beta }{\chi _{1}^{-}+\gamma }%
\end{bmatrix}%
,\quad \,M_{1}=\mathrm{diag}\left( \mathrm{e}^{\omega _{1}^{+}},\mathrm{e}%
^{\omega _{1}^{-}}\right) ,\,\,
\end{equation*}%
with
\begin{equation*}
\omega _{1}^{\pm }=\frac{\mathrm{i}}{4}(\chi _{1}^{\pm }-\lambda _{1})\left(
s+\frac{2}{\lambda _{1}}y\right) \pm a_{1},\text{ \ }\chi _{1}^{\pm
}=\lambda _{1}\pm \sqrt{(\lambda _{1}+\gamma )^{2}-\beta ^{2}}.
\end{equation*}
However, the soliton solution obtained above is usually singular. In order
to derive the one-dark soliton solution through the DT method, a limit
process $\lambda _{1}\rightarrow \bar{\lambda }_{1}$ is needed. To this
end, we first pick up one special solution
\begin{equation*}
|y_{1}\rangle =KL_{1}M_{1}%
\begin{bmatrix}
1 \\[8pt]
\alpha _{1}(\bar{\lambda}_{1}-\lambda _{1})%
\end{bmatrix}\,,%
\end{equation*}
further, for the sake of convenience, we set
\begin{equation*}
\lambda _{1}=\beta \cosh (\epsilon +\mathrm{i}\varphi _{1})-\gamma ,\quad \,\chi
_{1}^{\pm }=\beta \mathrm{e}^{\pm (\epsilon +\mathrm{i}\varphi _{1})}-\gamma,\quad \alpha _{1}=-\frac{\mathrm{e}^{-\mathrm{i}\varphi _{1}}}{4\beta \sin
^{2}\varphi _{1}}
\end{equation*}
where $\varphi _{1}\in (0,\pi )$.  By taking a limit $\epsilon \rightarrow 0$, we can obtain
\begin{equation}
\frac{\langle y_{1}|\sigma _{3}|y_{1}\rangle }{2(\bar{\lambda}_{1}-\lambda
_{1})}=\frac{\mathrm{e}^{2\omega _{1}}+1}{\beta (\mathrm{e}^{-\mathrm{i}%
\varphi _{1}}-\mathrm{e}^{\mathrm{i}\varphi _{1}})}
\end{equation}
where
\begin{equation*}
\omega _{1}=-\frac{\beta \sin \varphi_{1}}{4}\left( s+\frac{2y}{\beta
\cos \varphi_{1}-\gamma }\right) +a_{1}\,.
\end{equation*}
Thus, the single dark soliton can be written as
\begin{equation}
q [1]= \frac{\beta }{2}\left[ \frac{1+\mathrm{e}^{2(\omega _{1}-\mathrm{i}%
\varphi _{1})}}{1+\mathrm{e}^{2\omega _{1}}}\right] \mathrm{e}^{\mathrm{i}%
\theta}= \frac{\beta }{4}\left((1+\mathrm{e}^{-2\mathrm{i}\varphi _{1}}) + (\mathrm{e}^{-2\mathrm{i}\varphi _{1}}-1) \tanh \omega _{1}  \right)\mathrm{e}^{\mathrm{i}%
\theta},
\end{equation}
\begin{equation}
x=-\frac{\gamma }{2} y+\frac{\beta ^{2}}{8}s+\frac{\beta \sin \varphi _{1}%
\mathrm{e}^{2\omega _{1}}}{1+e^{2\omega _{1}}},\quad \,t=-s,
\end{equation}
The non-singularity condition for the single dark soliton is $\rho \lbrack
1]\neq 0$ for all $(x,t)\in \mathbb{R}^{2}$.  To analyze the property for the
one-soliton solution, we calculate out
\begin{equation}
\rho [1]=\frac{\partial x}{\partial y}=-\frac{\gamma e^{4\omega _{1}}+\left( 2+\frac{%
2\beta ^{2}\sin ^{2}\varphi_{1}}{\beta \cos \varphi_{1}-\gamma }%
\right) e^{2\omega _{1}}+\gamma }{2(1+e^{2\omega _{1}})^{2}}\,,
\end{equation}
thus we can classify this one-dark soliton solution as follows:
\begin{itemize}
\item \textbf{smooth soliton:} when $\gamma -\beta \cos \varphi_{1}<0$, or $\gamma -\beta \cos (\varphi
_{1})>0$ and $\Delta _{1}>0$, where $\Delta _{1}=2\gamma ^{2}-2\gamma \beta \cos \varphi_{1}-\beta
^{2}\sin ^{2}\varphi_{1}$, the single dark soliton solution is always smooth.
An example is illustrated in Fig. 2 (a).
\item \textbf{cuspon soliton:} when $\gamma -\beta \cos \varphi_{1}>0$ and $\Delta _{1}=0$
then $\rho [1]$ attains zero at only one point, which leads to
a cusponed dark soliton as displayed in Fig. 2 (b).

\item \textbf{loop soliton:} when
$\gamma -\beta \cos \varphi_{1}>0$,\,\,$\Delta _{1}<0$,
then $\rho [1]$ attains two zeros, which leads to a looped dark soliton.
\end{itemize}
The velocity can be solved with the following relation
\begin{equation*}
x-v_{sp,1}t-c_{1}=\frac{\gamma (\beta \cos \varphi_{1}-\gamma )}{\beta
\sin \varphi_{1}}\omega _{1}+\frac{\beta }{2}\sin \varphi_{1}\tanh
\left( \omega _{1}\right) .
\end{equation*}%
The velocity of dark soliton is
\begin{equation*}
v_{sp,1}=-\frac{2\gamma (\beta \cos \varphi_{1}-\gamma )+\beta ^{2}}{8},
\end{equation*}%
and the initial center is
\begin{equation*}
c_{1}=\frac{\gamma (\gamma -\beta \cos \varphi_{1})}{\beta \sin (\varphi
_{1})}a_{1}+\frac{\beta }{2}\sin \varphi_{1}.
\end{equation*}%
The trough of the dark soliton $|q|$ is along the line
\begin{equation*}
x-v_{sp,1}t-c_{1}=0\,,
\end{equation*}%
and the depth of the trough is $\frac 12 |{\beta} (1-\cos \varphi_{1})|$.
\begin{figure}[tbh]
\centering
\subfigure[$|q|$]{%
\includegraphics[height=50mm,width=70mm]{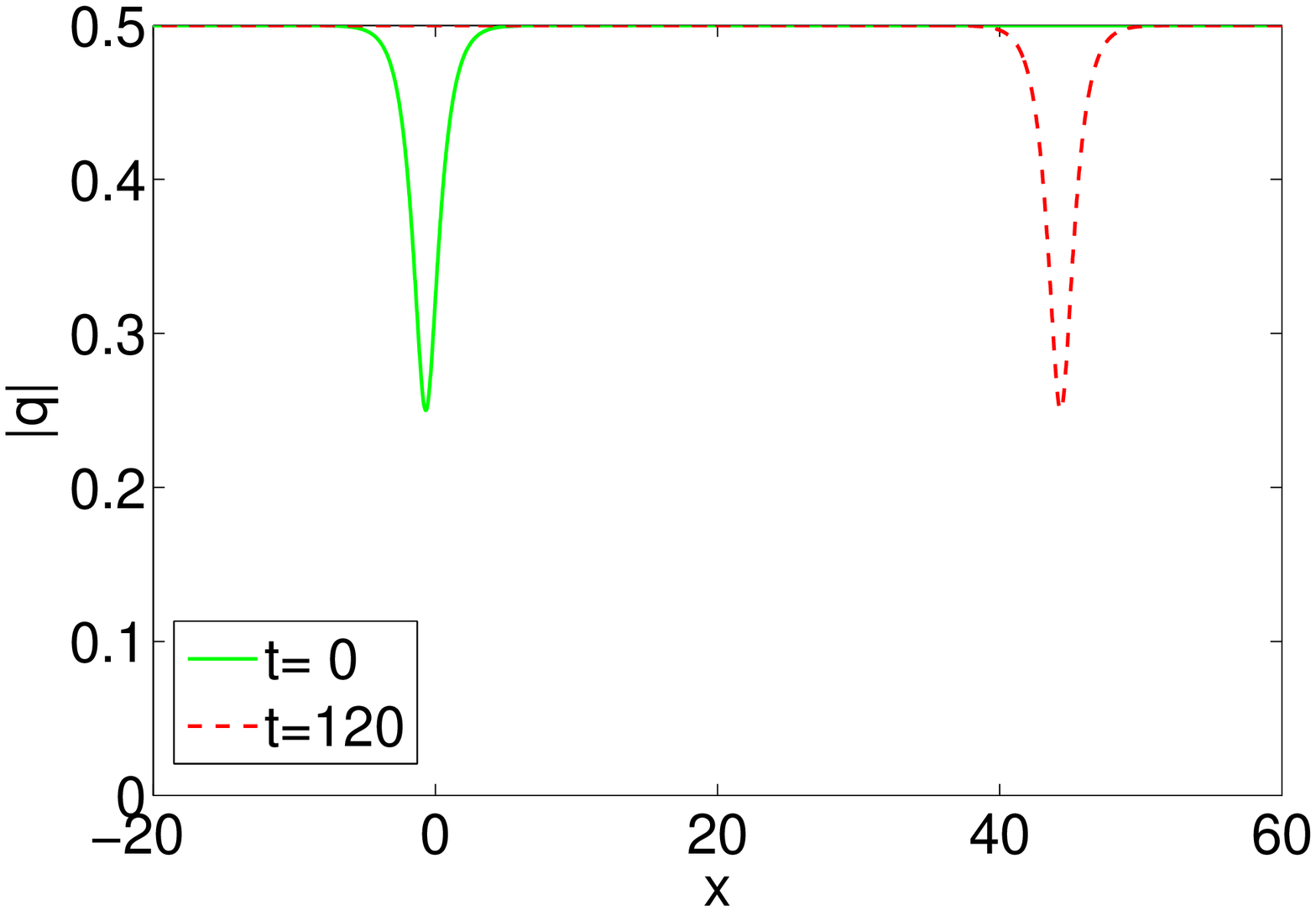}} \hfil
\subfigure[$|q|$]{%
\includegraphics[height=50mm,width=70mm]{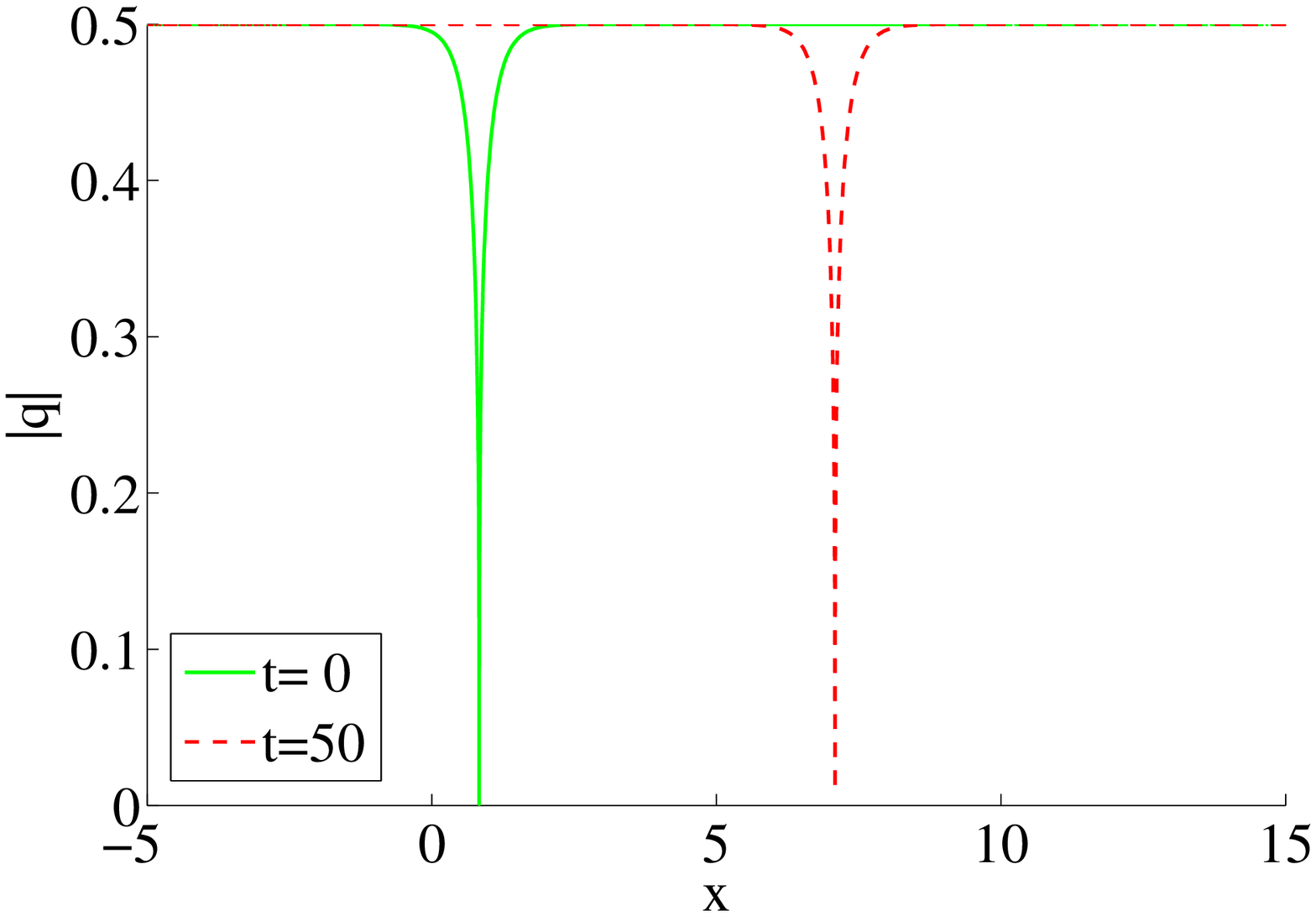}}
\caption{(color online): (a) Smooth dark soliton ($t=0$: blue solid line, $t=120$: red dashed line). Parameters $\protect\beta %
=1$, $\protect\gamma =1$, $\protect\varphi _{1}=2\protect\pi /3$, $a_{1}=0$;
(b) Cuspon dark soliton ($t=0$: blue solid line,  $t=50$: red dashed line). Parameters $\protect\beta =1$, $\protect\gamma =%
\protect\sqrt{2}/2$, $\protect\varphi _{1}=\protect\pi /2$, $a_{1}=0$; here this cuspon solution is a stationary solution. }
\label{fig3}
\end{figure}
\subsection{Multi-dark soliton solution}
Similar to the process of obtaining the single dark soliton solution,
starting with the same seed solution, and solving the Lax pair equation %
\eqref{ccd-laxa}--\eqref{ccd-laxb} with $(q=q[0],\rho =\rho
\lbrack 0])$ at $\lambda =\lambda
_{i}$, we have
\begin{equation*}
|y_{i}\rangle =KL_{i}M_{i}%
\begin{bmatrix}
1 \\[8pt]
\alpha _{i}(\bar{\lambda}_{i}-\lambda _{i})%
\end{bmatrix}%
\equiv K%
\begin{bmatrix}
\widehat{\phi _{i}} \\[8pt]
\beta \widehat{\psi _{i}}%
\end{bmatrix}%
\end{equation*}%
where
\begin{equation*}
\,\,L_{i}=%
\begin{bmatrix}
1 & 1 \\[8pt]
\frac{\beta }{\chi _{i}^{+}+\gamma } & \frac{\beta }{\chi _{i}^{-}+\gamma }%
\end{bmatrix}%
,\quad \,M_{i}=\mathrm{diag}\left( \mathrm{e}^{\omega _{i}^{+}},\mathrm{e}%
^{\omega _{i}^{-}}\right) ,\,\,
\end{equation*}
with
\begin{equation*}
\omega _{i}^{\pm }=\frac{\mathrm{i}}{4}(\chi _{i}^{\pm }-\lambda _{i})\left(
s+\frac{2}{\lambda _{i}}y\right) \pm a_{i},\quad \chi _{i}^{\pm
}=\lambda _{i}\pm \sqrt{(\lambda _{i}+\gamma )^{2}-\beta ^{2}},
\end{equation*}
$\alpha_i$s are appropriate complex parameters and $a_i$s are real parameters.
Based on the $N$-soliton solution (\ref{CSP-formula1})--(\ref{CSP-formula2}) to the defocusing CSP equation, it then follows
\begin{equation}
\begin{split}
q[N]=& \frac{\beta }{2}\left[ 1+\widehat{Y_{2}}M^{-1}\widehat{Y_{1}}^{\dag }%
\right] \mathrm{e}^{\mathrm{i}\theta }=\frac{\beta }{2}\left[ \frac{\det (H)%
}{\det (M)}\right] \mathrm{e}^{\mathrm{i}\theta }, \\
x=& -\frac{\gamma }{2}y+\frac{\beta ^{2}}{8}s-2\ln _{s}(\det (M)),\quad \,t=-s,
\end{split}
\label{nsoliton}
\end{equation}
where
\begin{equation*}
\begin{split}
M& =\left( \frac{\langle y_{i}|\sigma _{3}|y_{j}\rangle }{2(\bar{\lambda}_{i}-\lambda _{j})}\right)_{1\leq i,j\leq N},\quad \,H=M+Y_{1}^{\dag
}Y_{2}, \\
\widehat{Y_{1}}& =\left[ \widehat{\phi _{1}},\widehat{\phi _{2}},\cdots ,%
\widehat{\phi _{N}}\right] ,\quad \,\widehat{Y_{2}}=\left[ \widehat{\psi _{1}},%
\widehat{\psi _{2}},\cdots ,\widehat{\psi _{N}}\right] .
\end{split}%
\end{equation*}
In general, the above $N$-soliton solution \eqref{nsoliton} is singular.
In order to derive the $N$-dark soliton solution through the DT
method, we need to take a limit process $\lambda _{i}\rightarrow \bar{%
\lambda }_{i},(i=1,2,\cdots ,N)$ similar to the one in \cite{ling-dark}.
By a tedious procedure which is omitted here, we finally have the
$N$-dark soliton solution to the defocusing CSP equation
(\ref{dCSP}) as follows
\begin{equation}
q =\frac{\beta }{2}\frac{\det (H)%
}{\det (M)}\mathrm{e}^{\mathrm{i}\left(y+\frac{\gamma }{2}s\right)}\,,
\label{ndark1}
\end{equation}
\begin{equation}
x= -\frac{\gamma }{2}y+\frac{\beta ^{2}}{8}s-2\ln _{s}(\det (M)),\quad \,t=-s\,.
\label{ndark2}
\end{equation}
where the entries of the matrices $M$ and $H$  are
\begin{equation}
m_{i,j}=\frac{\mathrm{e}^{\omega _{i}+\omega _{j}}+\delta _{i,j}%
}{\beta (\mathrm{e}^{-\mathrm{i}\varphi _{i}}-\mathrm{e}^{\mathrm{i}\varphi
_{j}})}\,,\quad
h_{i,j}=\frac{\mathrm{e}^{(\omega _{i}-\mathrm{i}\varphi
_{i})+(\omega _{j}-\mathrm{i}\varphi _{j})}+\delta _{i,j}}{\beta (\mathrm{e}%
^{-\mathrm{i}\varphi _{i}}-\mathrm{e}^{\mathrm{i}\varphi _{j}})}\,\,, 1\leq i,j\leq N\,,
\label{n-dark-entry}
\end{equation}
$\delta_{i,j}$ is a Kronecker's delta and
\begin{equation}
\omega _{i}=-\frac{\beta }{4}\sin \varphi _{i}\left( s+\frac{2y}{\beta
\cos \varphi _{i}-\gamma }\right) +a_{i},\,\, \varphi_i\in\mathbf{R}.
\end{equation}%
By taking $N=2$ in (\ref{n-dark-entry}),
the determinants corresponding to two-dark soliton solution can be
calculated as
\begin{equation}
|M|=1+\mathrm{e}^{2\omega _{1}}+\mathrm{e}^{2\omega _{2}}+ a_{12} \mathrm{e}^{2(\omega _{1}+\omega _{2})}\,,
\end{equation}
\begin{equation}
|H|=1+\mathrm{e}^{2(\omega _{1}-\mathrm{i}\varphi _{1})}+\mathrm{e}^{2(\omega _{2}-\mathrm{i}%
\varphi _{2})}+a_{12}
\mathrm{e}^{2(\omega _{1}+\omega _{2}-\mathrm{i}\varphi _{1}-\mathrm{i}\varphi _{2})}\,,
\end{equation}
where
\begin{equation}
a_{12} =\frac{\sin ^{2}\left( \frac{\varphi
_{2}-\varphi _{1}}{2}\right) }{\sin ^{2}\left( \frac{\varphi _{2}+\varphi
_{1}}{2}\right) }\,.
\end{equation}
The collision processes between smooth-smooth dark solitons and smooth-cuspon dark solitons are illustrated in Figs. 3 (a) and (b), respectively.
It is seen that the interactions between dark solitons are elastic.
Different from the interaction between two smooth bright soliton, which could develop
singularity, the interaction between two smooth dark soliton never appears singularity. When a smoothed dark soliton interacts with a cusponed dark soliton, the singularity of the cusponed dark soliton could vanish as observed from the second figure in Fig 3(b).

\begin{figure}[tbh]
\centering
\subfigure[$|q|$]{%
\includegraphics[height=50mm,width=70mm]{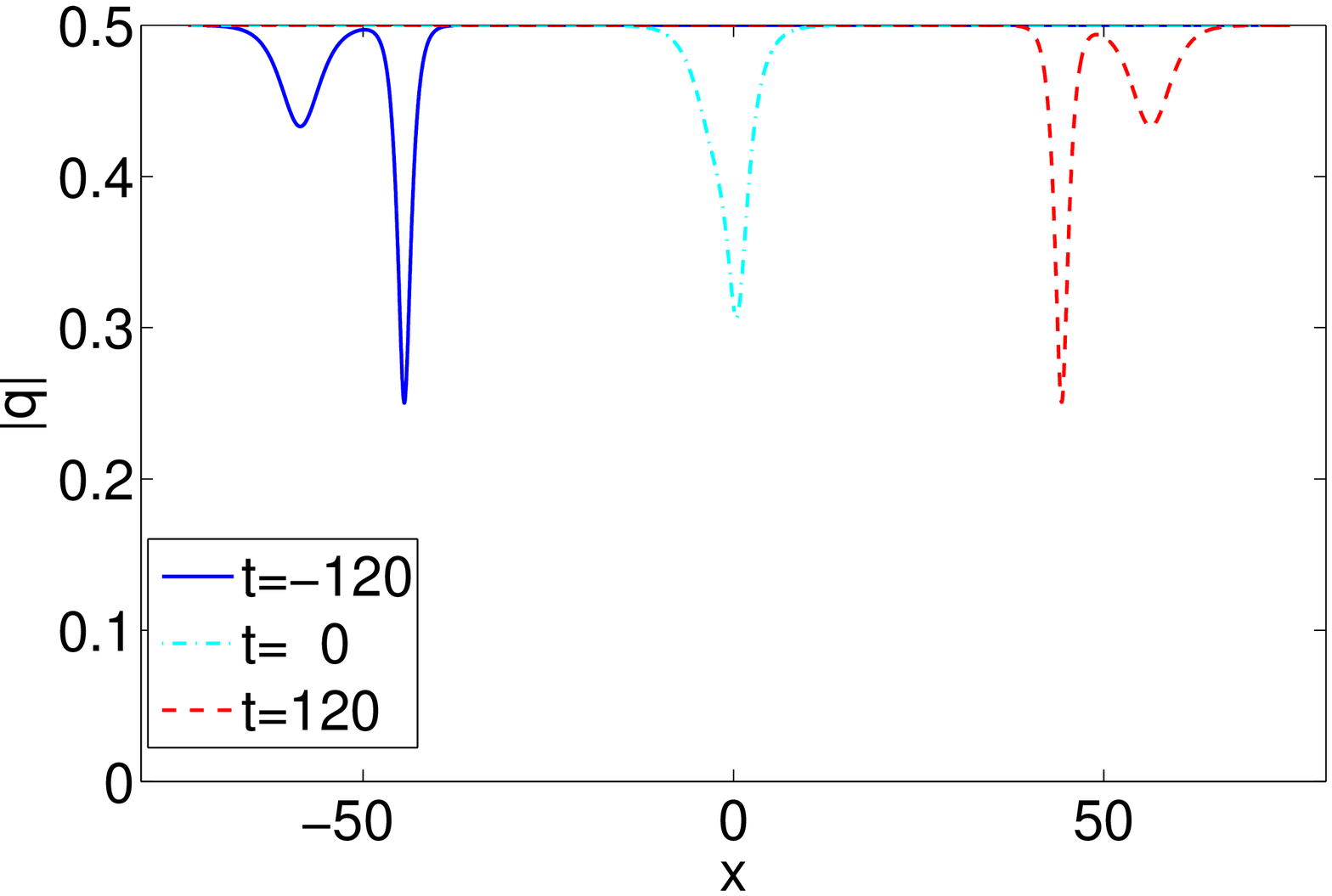}} \hfil
\subfigure[$|q|$]{%
\includegraphics[height=50mm,width=70mm]{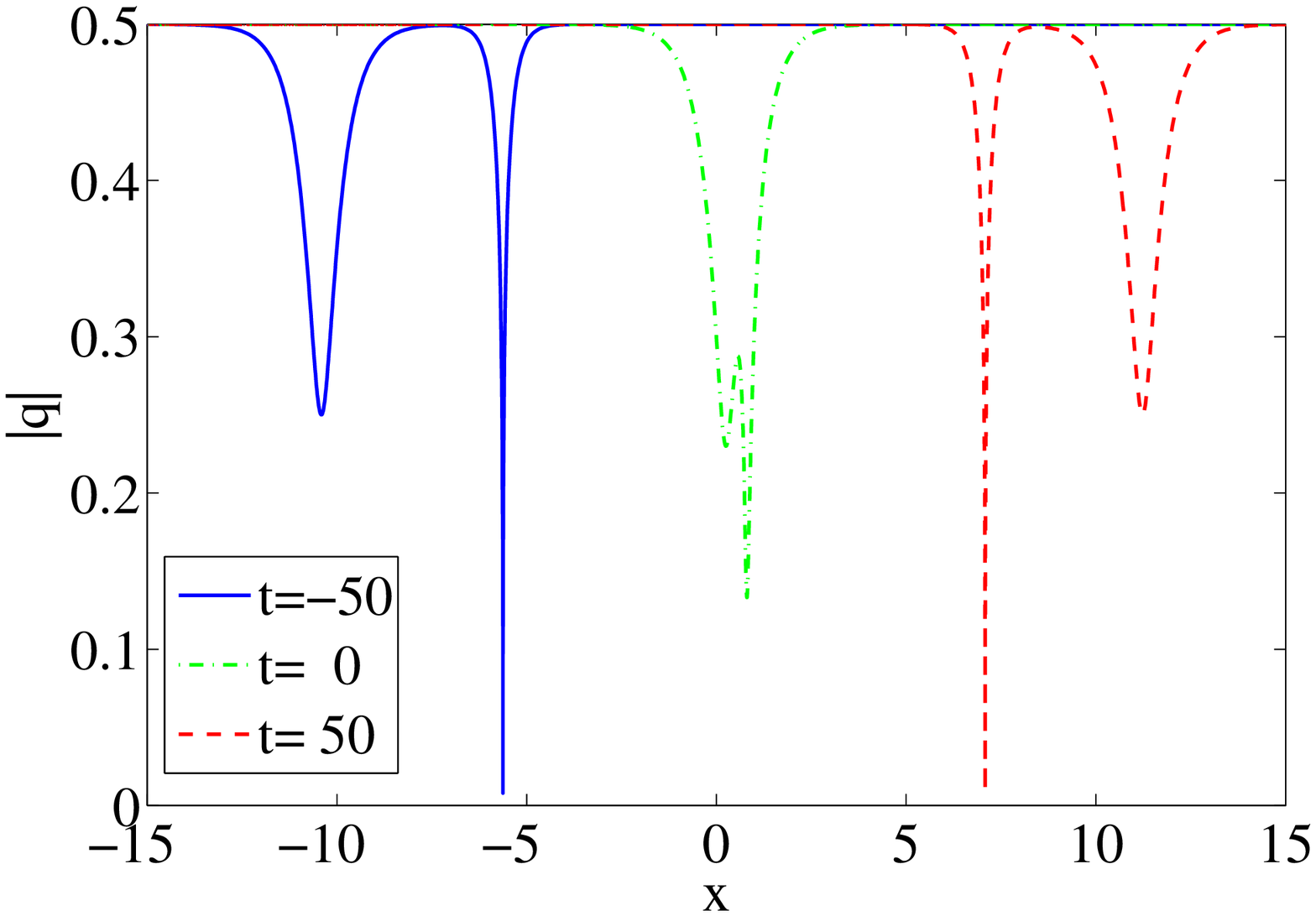}}
\caption{(color online): (a) Smooth-smooth dark soliton ($t=-120$: blue solid line, $t=0$: magenta dash-dot line, $t=120$: red dashed line). Parameters $\protect\beta %
=1$, $\protect\gamma =1$, $\protect\varphi_{1}=2\protect\pi /3$,  $\protect\varphi_{2}=5\protect\pi /6$, $a_{1}=a_{2}=0$;
(b) Smooth-cuspon dark soliton ($t=-50$: blue solid line, $t=0$: green dash-dot line, $t=50$: red dashed line). Parameters $\protect\beta =1$, $\protect\gamma =%
\protect\sqrt{2}/2$, $\protect\varphi _{1}=\protect\pi /2$, $\varphi _{2}=\protect2\pi /3$ $a_{1}=a_{2}=0$; It is
seen that the interaction of smooth dark soliton and cuspon dark soliton is elastic. }
\label{fig4}
\end{figure}

\section{Conclusions and discussions}
In this paper, we derived the complex short pulse equation of both focusing and defocusing types from the context of nonlinear optics and found the multi-dark soliton solution of the defocusing type. Comparing with the classical theory for the SP equation, there are several advantages in using complex representation. Firstly, amplitude and
phase are two fundamental characteristics for a wave packet, the information
of these two factors are nicely combined into a single complex-valued
function. Secondly, the use of complex representation can allow us to model the propagation of
optical pules in both the focusing and defocusing nonlinear media.
Such advantages can be observed in many analytical results related to the NLS equation and the CSP equation.
Therefore, by using a complex representation, we have shown that the focusing CSP equation admits
the bright soliton solution \cite{Feng_ComplexSPE,FengShen_ComplexSPE}, the breather solution as well as
the rogue wave solution \cite{LingFeng1}. Whereas, as shown in the present paper, the defocusing CSP equation has the
 multi-dark soliton solution same as the defocusing NLS equation. It would be an very interesting topic to compare the
properties of ultra-short optical pulses experimentally with the theoretical predictions for the CSP
 equation and the ones for the NLS equations. This, of course, is beyond the scope of the present paper.

The dynamics of dark soliton has been a hot topic in nonlinear optics. The history for the observation of dark soliton is even earlier than the bright soliton. In this work, we proposed a new integrable equation which admits multi-dark soliton solution. Moreover, we provided the dynamics analysis for the single dark soliton and two-dark soliton in detail. Especially, the cusponed dark soliton is found for the first time. The results would further enrich our understanding of dark solitons in the ultra-short pulse model.


\section*{Acknowledgment}
We are grateful to the anonymous referees for their constructive comments which help us to improve the original manuscript.
The work of BF is partially supported by the National Natural Science Foundation of China under grant 11428102, that of LML by National Natural Science Foundation of
China (Contact No. 11401221), that of ZNZ by the NSFC under grant 11271254,
 and in part by the Ministry of
Economy and Competitiveness of Spain under contract MTM2012-37070.

\end{document}